\documentclass[a4paper,11pt]{article}
\pdfoutput=1
\usepackage{jcappub}
\usepackage{afterpage,booktabs}
\usepackage{graphicx,subfig}
\usepackage{amssymb,amsmath,bm}
\usepackage[usenames,dvipsnames]{xcolor}
\usepackage[capitalise]{cleveref}
\usepackage{expl3,xstring}
\usepackage{dcolumn}
\usepackage{bm}
\usepackage{color}
\usepackage{breqn}

\newcommand{\copter}{\texttt{Copter}}
\newcommand{\MGcopter}{\texttt{MG-Copter}}
\newcommand{\cola}{\texttt{COLA}}
\newcommand{\picola}{\texttt{L-PICOLA}}
\newcommand{\MGpicola}{\texttt{MG-PICOLA}}
\newcommand{\frz}{$|f_{R0}|$}

\title{\boldmath Investigating the degeneracy between modified gravity and massive neutrinos with redshift-space distortions}
\author{Bill S. Wright$^1$,}
\author{Kazuya Koyama$^1$,}
\author{Hans A. Winther$^1$,}
\author{and Gong-Bo Zhao$^{2,3,1}$}
\affiliation{$^1$Institute of Cosmology \& Gravitation, University of Portsmouth, Portsmouth, Hampshire, PO1 3FX, UK}
\affiliation{$^2$National Astronomy Observatories, Chinese Academy of Science, Beijing, 100101, P.R.China}
\affiliation{$^3$University of Chinese Academy of Sciences, Beijing, 100049, P.R.China}
\emailAdd{bill.wright@port.ac.uk}

\date{\today}

\abstract{There is a well known degeneracy between the enhancement of the growth of large-scale structure produced by modified gravity models and the suppression due to the free-streaming of massive neutrinos at late times. This makes the matter power-spectrum alone a poor probe to distinguish between modified gravity and the concordance $\Lambda$CDM model when neutrino masses are not strongly constrained. In this work, we investigate the potential of using redshift-space distortions (RSD) to break this degeneracy when the modification to gravity is scale-dependent in the form of Hu-Sawicki $f(R)$. We find that if the linear growth rate can be recovered from the RSD signal, the degeneracy can be broken at the level of the dark matter field. However, this requires accurate modelling of the non-linearities in the RSD signal, and we here present an extension of the standard perturbation theory based model for non-linear RSD that includes both Hu-Sawicki $f(R)$ modified gravity and massive neutrinos.}

\begin{document}
\maketitle

\section{Introduction}

Modifications to Einstein's theory of General Relativity (GR) can be considered when searching for an explanation of the late-time acceleration \cite{KoyamaMGReview, CliftonMGReview}. The simplest class of models, known as scalar-tensor theories \cite{Horndeski1,Horndeski2,Horndeski3}, introduce a new scalar field that causes the required acceleration. This new scalar field also couples to matter, leading to a so-called fifth force. Such models typically invoke a screening mechanism to ensure that the fifth force would become negligible in high density environments, and thus not be observable in solar system tests. However, in other environments the fifth force is present and can enhance structure formation. This enhancement can be used to constrain such scalar-tensor theories of modified gravity (MG) with large-scale structure observations, and this is indeed a key goal of upcoming large-scale structure surveys such as Euclid \cite{Euclid}, DESI \cite{DESI}, WFIRST \cite{WFIRST}, LSST \cite{LSST}, and SKA \cite{SKA}.

In this paper we will consider $f(R)$ gravity \cite{fofr1, fofr2}, and in particular the commonly studied Hu-Sawicki variant of $f(R)$ \cite{HuSawicki}. This model has been well studied with {\it N}-body simulations as it is well understood theoretically and offers a phenomenology that is representative of a broad range of modified gravity models.

In GR with cold dark matter the growth rate of matter perturbations is scale-independent. A key signature of modified gravity is that the linear growth rate can be scale-dependent. However, a vital but often overlooked complication when searching for signatures of modified gravity in large-scale structure is the suppression of structure growth due to massive neutrinos. Neutrinos were first shown to have mass in observations of neutrino flavour oscillations \cite{MassiveNeutrinos1, MassiveNeutrinos2}, the presence of which demand that at least two of the neutrino states are massive \cite{Pontecorvo1957}. Even though particle physics experiments do not yet tell us the absolute mass of each of the three mass eigenstates, they do allow strong constraints to be placed on the difference in mass between the states, and these imply that at least one of the mass eigenstates has a mass $m_{\nu}\gtrsim 0.05~\mathrm{eV}$ \cite{PDG2018}. As a consequence of having mass, the matter-radiation equality time will be delayed and the neutrinos will not cluster at scales below their free-streaming length $\lambda_{\rm fs}$ \cite{MassiveNeutrinoReview}. The delay to the time of matter-radiation equality lowers the amplitude of the matter perturbations at the start of matter-domination, and the free-streaming of massive neutrinos causes the dark matter perturbations to feel a reduced gravitational potential below $\lambda_{\rm fs}$ and thus cluster less strongly than in a model with the same value of the matter density parameter but only massless neutrinos. The combination of these two effects leads to a scale-dependent suppression of structure growth; a signature which can be used to constrain the neutrino masses if it can be measured by the previously mentioned large-scale structure surveys \cite{Bondetal1980,2016PhRvD..94h3522G,2016PDU....13...77C,2015JCAP...02..045P,2014MNRAS.444.3501B,2013MNRAS.436.2038Z,2017PhRvD..96l3503V,2018PhRvD..98l3526G}, even in models beyond $\Lambda$CDM that affect structure growth in a scale-independent way \cite{BoyleKomatsu2018}.

However, with the potential for scale-dependent enhancement of structure formation from modified gravity and the scale-dependent suppression due to massive neutrinos, there is a risk of degeneracy whereby large-scale structure in a universe with a strong modification to gravity and heavy neutrinos can be difficult to distinguish from that of a universe with GR and light neutrinos \cite{2013PhRvL.110l1302M,2013PhRvD..88j3523H,Baldi2014,2015PhRvD..91f3524H,2010PThPh.124..541M,Bellomo2017,2017PhRvD..95f3502A}. This degrades the ability of surveys to achieve their twin goals of testing gravity and constraining the neutrino masses in any theories of gravity beyond GR. Indeed, it has been shown that the non-linear matter power spectrum \cite{Baldi2014} and halo mass function \cite{Hagstotzetal2018} in $f(R)$ models are difficult to distinguish from their equivalents in GR when the neutrino masses are allowed to vary. The DES Collaboration considers neutrino mass and extensions to GR in the same analysis \cite{DESjoint}, although they only state the resulting constraints on the MG parameters and not the neutrino masses. There are some promising signs that certain observables may be better at reducing or even breaking this degeneracy, such as higher-order weak lensing statistics \cite{Peeletal2018} and weak lensing tomographic information at multiple redshifts \cite{Giocolietal2018}; as well as techniques that are superior at distinguishing models such as machine learning \cite{DegenMachineLearning1,DegenMachineLearning2}.

A different observable that has degeneracy breaking potential is that of redshift-space distortions (RSD) \cite{RSDReview}. RSD occur when the distances to tracers are computed using their observed redshifts without accounting for the effect of the tracers' peculiar velocities on the redshifts which adds to the contribution from the Hubble flow. On linear scales this results in a slight squashing along the line-of-sight \cite{Kaiser1987}, whereas there is a strong stretching along the line-of-sight at non-linear scales commonly known as the Fingers-of-God (FoG) effect \cite{FoG}. For combinations of MG parameters and neutrino masses whose enhancement and suppression of {\it structure growth} produce matter power spectra that are difficult to differentiate between, the {\it structure growth rate} can still be different in each case and allow for models to be distinguished between. It has recently been shown that growth rate information imprinted in velocity statistics in real-space can be used to break the degeneracy \cite{Hagstotzetal2019}. However, real-space velocity statistics are not directly observable. Fortunately, because of the velocity information encoded in them, RSD observations can be used to extract the linear growth rate of structure $f$. However, in order to extract $f$ and break the degeneracy, it is necessary to accurately model the non-linearities of RSD with MG and massive neutrinos.

In this paper, we extend the cosmological perturbation theory code \copter{} \cite{OriginalCOPTER} to include the effects of massive neutrinos in addition to those of modified gravity allowing us to accurately model non-linear RSD in scenarios with Hu-Sawicki $f(R)$ gravity and non-zero neutrino masses. We build on \MGcopter{}, the modified version of \copter{} developed in \cite{BoseKoyama2016}, which is itself based on the approach presented in \cite{Taruya}.

We validate this implementation against simulations using the COmoving Lagrangian Acceleration (COLA) method \cite{OriginalCOLA}, which is a fast approximate simulation method, and then investigate whether the degeneracy between the two effects is broken by RSD at the level of the dark matter field.

The paper is organised as follows. In Section~\ref{sec:Implementation} we explain our implementation of modified gravity and massive neutrinos in the Standard Perturbation Theory (SPT) formalism and \MGcopter{} code. In Section~\ref{sec:Validation}, we show the results of tests validating our implementation against simulation results. In Section~\ref{sec:Degeneracy} we use our new implementation to investigate the degeneracy and then conclude in Section~\ref{sec:Conclusion}.

\section{Implementation} \label{sec:Implementation}

In order to model the combined effect of modified gravity and massive neutrinos on real- and redshift-space power spectra with low computational expense, it is necessary to include both effects in a semi-analytical code such as \copter{} which computes large-scale structure observables using perturbation theory. For the redshift-space quantities, \copter{} depends on the TNS model of redshift-space distortions which is named after the authors of \cite{TNS} (Taruya, Nishimichi, and Saito).

\subsection{\MGcopter{} and the TNS model} \label{ssec:LCDM_imp}
\MGcopter{} \cite{BoseKoyama2016} solves the equations of Standard Perturbation Theory (SPT) to acquire the real-space power spectra up to 1-loop order based on the approach developed by \cite{Taruya}.

Starting from the continuity and Euler equations, assuming fluid quantities to be irrotational such that velocity field $\Vec{v}$ can be expressed in terms of the velocity divergence $\theta=\left( \Vec{\nabla} \cdot \Vec{v} \right)/aH$, and transforming to Fourier space yields
\begin{align}
    a \frac{\partial\delta(\Vec{k})}{\partial a} + \theta(\Vec{k})
    &= -\int \frac{d^3\Vec{k}_1 d^3\Vec{k}_2}{(2\pi)^3} \delta_D\left( \Vec{k} - \Vec{k}_1 - \Vec{k}_2 \right) \alpha(\Vec{k}_1, \Vec{k}_2) \theta(\Vec{k}_1)\delta(\Vec{k}_2)~, \label{eq:continuity}
    \\
    \nonumber \\
    a \frac{\partial\theta(\Vec{k})}{\partial a} + \left( 2 + \frac{aH^{\prime}}{H} \right)\theta(\Vec{k}) &- \left( \frac{k}{aH} \right)^2 \Phi(\Vec{k})
    \nonumber \\ &= -\frac{1}{2}\int \frac{d^3\Vec{k}_1 d^3\Vec{k}_2}{(2\pi)^3} \delta_D\left( \Vec{k} - \Vec{k}_1 - \Vec{k}_2 \right) \beta(\Vec{k}_1, \Vec{k}_2) \theta(\Vec{k}_1)\theta(\Vec{k}_2)~, \label{eq:Euler}
\end{align}
where $a=1/(1+z)$ is the scale factor, $y^{\prime}=\partial y/\partial a$ and the kernels $\alpha$ and $\beta$ are given by
\begin{align}
    \alpha(\Vec{k}_1, \Vec{k}_2) &= 1 + \frac{\Vec{k}_1 \cdot \Vec{k}_2}{|\Vec{k}_1|^2}~,
    \\
    \nonumber \\
    \beta(\Vec{k}_1, \Vec{k}_2) &= \frac{(\Vec{k}_1 \cdot \Vec{k}_2)|\Vec{k}_1+\Vec{k}_2|^2}{|\Vec{k}_1|^2|\Vec{k}_2|^2}~.
\end{align}
The Poisson equation completes the above modified continuity and Euler equations
\begin{align}
    -\left( \frac{k}{aH} \right)^2 \Phi(\Vec{k}) = \frac{3\Omega_{\rm m}(a)}{2}\delta(\Vec{k})~,
\end{align}
where $\Omega_{\rm m}(a) = 8\pi G\rho_{\rm m}/3H^2$. We want the $n^{\rm th}$ order solutions of \cref{eq:continuity,eq:Euler}
to be of the form
\begin{align}\label{eq:deltasol}
    \delta_n(\Vec{k}, a) = \int d^3\Vec{k}_1 \ldots d^3\Vec{k}_n \delta_D(\Vec{k}-\Vec{k}_{1 \ldots n})F_n(\Vec{k}_1,\ldots,\Vec{k}_n, a) \Delta(\Vec{k}_1)\ldots\Delta(\Vec{k}_n)~,
\end{align}
\begin{align}\label{eq:thetasol}
    \theta_n(\Vec{k}, a) = \int d^3\Vec{k}_1 \ldots d^3\Vec{k}_n \delta_D(\Vec{k}-\Vec{k}_{1 \ldots n})G_n(\Vec{k}_1,\ldots,\Vec{k}_n, a) \Delta(\Vec{k}_1)\ldots\Delta(\Vec{k}_n)~,
\end{align}
where $\Vec{k}_{1 \ldots n}=\Vec{k}_{1}+\ldots+\Vec{k}_n$. Inserting these forms of the solutions into \cref{eq:continuity,eq:Euler} yields a generalised system of equations for the $n^{\rm th}$ order kernels \cite{Taruya}
\begin{align}
    \hat{\mathcal{L}} \begin{bmatrix} F_n(\Vec{k}_1,\ldots,\Vec{k}_n) \\ G_n(\Vec{k}_1,\ldots,\Vec{k}_n) \end{bmatrix}
    = \sum_{j=1}^{n-1} \begin{bmatrix} -\alpha(\Vec{k}_{1 \ldots j}, \Vec{k}_{j+1 \ldots n}) G_j(\Vec{k}_1,\ldots,\Vec{k}_j) F_{n-j}(\Vec{k}_{j+1},\ldots,\Vec{k}_n) \\ -\frac{1}{2}\beta(\Vec{k}_{1 \ldots j}, \Vec{k}_{j+1 \ldots n}) G_j(\Vec{k}_1,\ldots,\Vec{k}_j) G_{n-j}(\Vec{k}_{j+1},\ldots,\Vec{k}_n) \end{bmatrix}~,
\end{align}
where
\begin{align}
    \hat{\mathcal{L}} = \begin{bmatrix} a\frac{d}{da} & 1 \\ \ \ \frac{3\Omega_{\rm m}}{2}\ \ \ \  & a\frac{d}{da} + \left( 2 + \frac{aH^{\prime}}{H} \right) \end{bmatrix}~.
\end{align}
\MGcopter{} solves this system of equations to compute the kernels $F_i$ and $G_i$. The power spectra up to 1-loop are given as
\begin{align}
    P^{\rm 1-loop}_{ij}(k) = P^{\rm L}_{ij}(k) + P_{ij}^{13}(k) + P_{ij}^{22}(k)~,
\end{align}
where the 1-loop corrections are defined by
\begin{align}
    \left\langle x_2(\Vec{k}) y_2(\Vec{k}^{\prime}) \right\rangle &= (2\pi)^3 \delta_D(\Vec{k}+\Vec{k}^{\prime}) P_{xy}^{22}(k)~,
    \nonumber \\
    \left\langle x_1(\Vec{k}) y_3(\Vec{k}^{\prime}) + x_3(\Vec{k}) y_1(\Vec{k}^{\prime}) \right\rangle &= (2\pi)^3 \delta_D(\Vec{k}+\Vec{k}^{\prime}) P_{xy}^{13}(k)~,
\end{align}
where $x$ and $y$ can be $\delta$ or $\theta$. Working these through, the final expressions for the 1-loop corrections in terms of the $z=0$ linear power spectrum $P_0(k)=P^{\rm L}(k, z=0)$ are, for the 22 correction,
\begin{align}
    P_{\delta\delta}^{22}(k) =& 2 \frac{k^3}{(2\pi)^2}\int_0^{\infty}r^2 \mathrm{d}r \int_{-1}^1 P_0(kr)P_0(k\sqrt{1+r^2-2rx})F_2^2(k, r, x) \mathrm{d}x~,
    \\
    \nonumber \\
    P_{\delta\theta}^{22}(k) =& 2 \frac{k^3}{(2\pi)^2}\int_0^{\infty}r^2 \mathrm{d}r \int_{-1}^1 P_0(kr)P_0(k\sqrt{1+r^2-2rx})F_2(k, r, x) G_2(k, r, x) \mathrm{d}x~,
    \\
    \nonumber \\
    P_{\theta\theta}^{22}(k) =& 2 \frac{k^3}{(2\pi)^2}\int_0^{\infty}r^2 \mathrm{d}r \int_{-1}^1 P_0(kr)P_0(k\sqrt{1+r^2-2rx})G_2^2(k, r, x) \mathrm{d}x~,
\end{align}
while for the 13 correction we have
\begin{align}
    P_{\delta\delta}^{13}(k) =& 6 \frac{k^3}{(2\pi)^2}F_1(k)P_0(k) \int_0^{\infty} r^2 P_0(kr)F_3(k, r, x) \mathrm{d}r~,
    \\
    \nonumber \\
    P_{\delta\theta}^{13}(k) =& 3 \frac{k^3}{(2\pi)^2}F_1(k)P_0(k) \int_0^{\infty} r^2 P_0(kr)G_3(k, r, x) \mathrm{d}r
    \nonumber \\ &+ 3 \frac{k^3}{(2\pi)^2}G_1(k)P_0(k) \int_0^{\infty} r^2 P_0(kr)F_3(k, r, x) dr~,
    \\
    \nonumber \\
    P_{\theta\theta}^{13}(k) =& 6 \frac{k^3}{(2\pi)^2}G_1(k)P_0(k) \int_0^{\infty} r^2 P_0(kr)G_3(k, r, x) \mathrm{d}r~.
\end{align}
With the SPT real-space power spectra computed up to 1-loop order, \MGcopter{} can then input these to the TNS model to calculate the redshift-space power spectrum $P^{(s)}(k)$.

The TNS model for the redshift-space power spectrum $P^{(s)}$ as a function of scale $k$ and line-of-sight (LoS) angle parameter $\mu=\cos(\theta)$ is given by Eq.~(18) of \cite{TNS}, which we reproduce here with subtle changes due to the different definition of $\theta$:
\begin{align} \label{eq:TNS}
    P^{(s)}(k, \mu) = D_{\rm FoG}\left[ k\mu \sigma_v \right] \left\{ P_{\delta\delta}(k) - 2\mu^2 P_{\delta\theta}(k) + \mu^4 P_{\theta\theta}(k) + A(k, \mu) + B(k, \mu) \right\}~,
\end{align}
where $D_{\rm FoG}$ is the Fingers-of-God damping function which we will discuss later. It is generally a function of $k$, $\mu$, and the velocity dispersion $\sigma_v$. The power spectra $P_{\delta\delta}(k)$, $P_{\delta\theta}(k)$, and $P_{\theta\theta}(k)$ correspond to the density auto-correlation, density-velocity divergence cross correlation, and the velocity divergence auto-correlation respectively. $A(k, \mu)$ and $B(k, \mu)$ are correction terms given by
\begin{align}
    A(k, \mu) &= - k\mu \int \frac{d^3 \Vec{p}}{(2\pi)^3}\frac{p_z}{p^2}
    \left\{ B_{\sigma}(\Vec{p}, \Vec{k} - \Vec{p}, -\Vec{k})-B_{\sigma}(\Vec{p}, \Vec{k}, -\Vec{k}-\Vec{p}) \right\}, \label{eq:A}
    \\
    B(k, \mu) &= (k\mu)^2 \int \frac{d^3 \Vec{p}}{(2\pi)^3} F(\Vec{p})F(\Vec{k}-\Vec{p})~, \label{eq:B}
\end{align}
where $B_{\sigma}$ is the cross bispectrum defined by
\begin{align}
    \left\langle \theta(\Vec{k}_1) \left\{ \delta(\Vec{k}_2) - \frac{k^2_{2z}}{k^2_2}\theta(\Vec{k}_2) \right\} \left\{ \delta(\Vec{k}_3) - \frac{k^2_{3z}}{k^2_3}\theta(\Vec{k}_3) \right\} \right\rangle
    \nonumber \\ = (2\pi)^3 \delta_D(\Vec{k}_1+\Vec{k}_2+\Vec{k}_3) B_{\sigma}(\Vec{k}_1,\Vec{k}_2,\Vec{k}_3)
    ~,
\end{align}
and $F(\Vec{p})$ is defined as
\begin{align}
    F(\Vec{p})=\frac{p_z}{p^2} \left\{ P_{\delta\theta}(p) - \frac{p_z^2}{p^2}P_{\theta\theta}(p) \right\}~.
\end{align}
Throughout we use an exponential form for the Fingers-of-God damping factor:
\begin{align}
    D_{\rm FoG}\left[ k\mu \sigma_v \right] = \exp\left( -k^2\mu^2 \sigma_v^2 \right)~.
\end{align}
The velocity dispersion $\sigma_v$ is a free parameter and needs to be fitted to some other $P^{(s)}$ data, for example from simulations as we do here. To do this, we minimise the likelihood function
\begin{align}
    -2\ln\mathcal{L} = \sum_n \sum_{l,l^{\prime}=0, 2} \left( P^{s}_{l,\ \copter{}}(k_n) - P^{s}_{l,\ \cola{}}(k_n) \right) \mathrm{Cov}^{-1}_{l,l^{\prime}}(k_n) \left( P^{s}_{l^{\prime},\ \copter{}}(k_n) - P^{s}_{l^{\prime},\ \cola{}}(k_n) \right)
\end{align}
for the first two multipoles. Expressions for the covariance matrix between the different multipoles $\mathrm{Cov}_{l,l^{\prime}}$ are given in Appendix C of \cite{TNS}. We do not consider non-Gaussianity in this covariance but we do include the effect of shot-noise. For the validation of our implementation of massive neutrinos in \MGcopter{} presented in Section~\ref{sec:Validation} we assume an ideal survey with survey volume $V_{\rm s}=10~{\rm Gpc}^3/h^3$ and galaxy number density $\bar{n}_{\rm g}=4\times 10^{-3} h^3/{\rm Mpc}$. For the study of the degeneracy in Section~\ref{sec:Degeneracy}, we want to model a slightly more realistic scenario, so we assume a DESI-like survey with $V_{\rm s}$ and $\bar{n}_{\rm g}$ as given in Table~\ref{tab:surv_param} and redshift bin width $\Delta z=0.2$. These values are computed using the information for emission line galaxies (ELGs) in Table~V of \cite{DESIparams}.

\begin{table}
\begin{center}
\begin{tabular}{ |c|c|c| }
 \hline
 $z$ & $V_{\rm s}$ (Gpc$^3$/$h^3$) & $\bar{n}_{\rm g}$ ($h^3$/Mpc$^3$) \\
 \hline
 0.5 & 3.40  & 2.95$\times 10^{-4}$ \\
 1.0 & 7.68  & 5.23$\times 10^{-4}$ \\
 1.5 & 10.14 & 1.71$\times 10^{-4}$ \\
 \hline
\end{tabular}
\caption{Survey parameters for a DESI-like survey computed from the information for emission line galaxies (ELGs) in Table~V of \cite{DESIparams}. These parameters are used in the computation of the covariance matrices for fitting $\sigma_v$ in \MGcopter{} in the study of the degeneracy in Section~\ref{sec:Degeneracy}.}
\label{tab:surv_param}
\end{center}
\end{table}

Thus the TNS model can be used to compute $P^{(s)}(k, \mu)$ with the input of $P_{\delta\delta}$, $P_{\delta\theta}$, $P_{\theta\theta}$ at 1-loop order from \MGcopter{}.

\subsection{Adding modified gravity} \label{ssec:MG_imp}

Modified gravity models, like the $f(R)$ gravity model we consider here, have been previously added to \copter{} in \cite{BoseKoyama2016}, resulting in \MGcopter{}. The 1-loop real-space power spectra are affected by the inclusion of modified gravity in SPT, but the TNS model of Eq.~(\ref{eq:TNS}) is still applicable without changes. We shall reproduce here the essentials of the implementation of modified gravity in the SPT part of \MGcopter{}.

The modifications to gravity can be included in the Poisson equation, which up to $3^{\rm rd}$ order becomes
\begin{align}\label{eq:MGPoisson}
    -\left( \frac{k}{aH} \right)^2 \Phi(\Vec{k}) = \frac{3\Omega_{\rm m}(a)}{2}\delta(\Vec{k})\mu(k, a) + S(\Vec{k}) ~,
\end{align}
where $\mu(k, a)=G_{\rm eff}(k, a)/G$ is an effective Newton's constant\footnote{Not to be confused with the line-of-sight angle parameter $\mu$, which will always be presented without arguments.} and the non-linear source term $S(\Vec{k})$ up to $3^{\rm rd}$ order is
\begin{align}
    S(\Vec{k}) = &\int \frac{d^3\Vec{k}_1 d^3\Vec{k}_2}{(2\pi)^3}\delta_D(\Vec{k}-\Vec{k}_{12})\gamma_2(\Vec{k}, \Vec{k}_1, \Vec{k}_2, a) \Delta(\Vec{k}_1)\Delta(\Vec{k}_2)
    \nonumber \\ &+ \int \frac{d^3\Vec{k}_1 d^3\Vec{k}_2 d^3\Vec{k}_3}{(2\pi)^3}\delta_D(\Vec{k}-\Vec{k}_{123})\gamma_3(\Vec{k}, \Vec{k}_1, \Vec{k}_2, \Vec{k}_3, a) \Delta(\Vec{k}_1)\Delta(\Vec{k}_2)\Delta(\Vec{k}_3)~.
\end{align}
While the effective Newton's constant $\mu(k, a)$ is generally responsible for the (scale-dependent) growth of linear perturbations, at the fully non-linear level modified gravity models typically include a screening mechanism that will affect the growth of non-linearities, and the $\gamma_{2}$ and $\gamma_{3}$ terms provide the leading order description of this screening in perturbation theory.

Using the same form for the $n^{\rm th}$ order solutions as in \cref{eq:deltasol,eq:thetasol}, the new system of equations for the $n^{\rm th}$ order kernels is
\begin{align}
    \hat{\mathcal{L}} \begin{bmatrix} F_n(\Vec{k}_1,\ldots,\Vec{k}_n) \\ G_n(\Vec{k}_1,\ldots,\Vec{k}_n) \end{bmatrix}
    = \sum_{j=1}^{n-1} \begin{bmatrix} -\alpha(\Vec{k}_{1 \ldots j}, \Vec{k}_{j+1 \ldots n}) G_j(\Vec{k}_1,\ldots,\Vec{k}_j) F_{n-j}(\Vec{k}_{j+1},\ldots,\Vec{k}_n) \\ -\frac{1}{2}\beta(\Vec{k}_{1 \ldots j}, \Vec{k}_{j+1 \ldots n}) G_j(\Vec{k}_1,\ldots,\Vec{k}_j) G_{n-j}(\Vec{k}_{j+1},\ldots,\Vec{k}_n) - N_n(\Vec{k}, \Vec{k}_1,\ldots,\Vec{k}_n) \end{bmatrix}~,
\end{align}
where
\begin{align}
    \hat{\mathcal{L}} = \begin{bmatrix} a\frac{d}{da} & 1 \\ \ \ \frac{3\Omega_{\rm m}}{2}\mu(k, a)\ \ \ \  & a\frac{d}{da} + \left( 2 + \frac{aH^{\prime}}{H} \right) \end{bmatrix}~,
\end{align}
and
\begin{align}
    N_2 =& \gamma_2(\Vec{k}, \Vec{k}_1, \Vec{k}_2)F_1(\Vec{k}_1)F_1(\Vec{k}_2)~,
    \\
    \nonumber \\
    N_3 = &\gamma_2(\Vec{k}, \Vec{k}_1, \Vec{k}_{23})F_1(\Vec{k}_1)F_2(\Vec{k}_2, \Vec{k}_3)
    + \gamma_2(\Vec{k}, \Vec{k}_{12}, \Vec{k}_3)F_2(\Vec{k}_1, \Vec{k}_2)F_1(\Vec{k}_3)
    \nonumber \\ &+ \gamma_3(\Vec{k}, \Vec{k}_1, \Vec{k}_2, \Vec{k}_3)F_1(\Vec{k}_1)F_1(\Vec{k}_2)F_1(\Vec{k}_3)~.
\end{align}

In this work we investigate Hu-Sawicki $f(R)$ gravity, which has a single free parameter $|f_{R0}|$. Hu-Sawicki $f(R)$ gravity produces an enhanced, scale-dependent growth of density perturbations relative to GR, but the built-in chameleon screening mechanism ensures that the modifications become negligible in high density environments (which typically coincide with small scales). Hu-Sawicki $f(R)$ only becomes active at late times, and thus the modifications to GR are negligible in the early Universe. For this theory, the extra terms in Eq.~(\ref{eq:MGPoisson}) are given as
\begin{align}
    \mu(k, a) =& ~1 + \left( \frac{k}{a} \right)^2 \frac{1}{3\Pi(k, a)}~,
    \\
    \nonumber \\
    \gamma_2(k, \Vec{k}_1, \Vec{k}_2, a) =& - \frac{9}{48} \left( \frac{kH_0}{aH} \right)^2 \left( \frac{H_0^2\Omega_{\mathrm{m}0}}{a^3} \right)^2
    \frac{(\Omega_{\mathrm{m}0}-4a^3(\Omega_{\mathrm{m}0}-1))^5}{a^{15}|f_{R0}|^2(3\Omega_{\mathrm{m}0}-4)^4}
    \nonumber \\ &\times\frac{1}{\Pi(k, a)\Pi(k_1, a)\Pi(k_2, a)}~,
    \\
    \nonumber \\
    \gamma_3(k, \Vec{k}_1, \Vec{k}_2, \Vec{k}_3, a) =& \left( \frac{kH_0}{aH} \right)^2 \left( \frac{H_0^2\Omega_{\mathrm{m}0}}{a^3} \right)^3
    \frac{1}{36\Pi(k, a)\Pi(k_1, a)\Pi(k_2, a)\Pi(k_3, a)\Pi(k_{23}, a)}
    \nonumber \\ &\times \left[ -\frac{45}{8}\frac{\Pi(k_{23}, a)}{a^{21}|f_{R0}|^3} \left(\frac{(\Omega_{\mathrm{m}0}-4a^3(\Omega_{\mathrm{m}0}-1))^7}{(3\Omega_{\mathrm{m}0}-4)^6}\right) \right.
    \nonumber \\ &\left. + H_0^2 \left( \frac{9}{4a^{15}|f_{R0}|^2} \frac{(\Omega_{\mathrm{m}0}-4a^3(\Omega_{\mathrm{m}0}-1))^5}{(3\Omega_{\mathrm{m}0}-4)^4} \right)^2 \right] ~,
\end{align}
where
\begin{align}
    \Pi(k, a) = \left( \frac{k}{a} \right)^2 + \frac{H_0^2 (\Omega_{\mathrm{m}0}-4a^3(\Omega_{\mathrm{m}0}-1))^3}{2|f_{R0}|a^9(3\Omega_{\mathrm{m}0}-4)^2}~.
\end{align}

\subsection{Adding massive neutrinos} \label{ssec:mass_nu_imp}

We have added support for massive neutrinos to the code \MGcopter{} developed in \cite{BoseKoyama2016}. Note that massive neutrinos were also added to the original \copter{} code in \cite{COPTERNeutrinos} using a similar approach. In our implementation, we follow the method of \cite{Saito:2008bp,Saito:2009ah} and include massive neutrinos at the level of the linear real-space power spectra $P^{\rm L}$, $P_{\delta\theta, \mathrm{L}}=f(k)P^{\rm L}$, and $P_{\theta\theta, \mathrm{L}}=f^2(k)P^{\rm L}$ without modifying the higher order SPT kernels. This allows us to take $P^{\rm L}(k)$ and $f(k)$ from \texttt{CAMB} \cite{CAMB} (or \texttt{MGCAMB} \cite{MGCAMB1, MGCAMB2} for MG+$m_{\nu}$) as input to \MGcopter{}; note that a small modification to \texttt{CAMB}/\texttt{MGCAMB} is necessary to get scale-dependent growth rate $f(k)$ as output. This method for including massive neutrinos is general enough to handle the various hierarchies of neutrino mass eigenstates \cite{NeutrinoMassHierarchy}, but for simplicity in the results that follow we have modelled the massive neutrinos as a single massive eigenstate with mass $m_{\nu}$ and two massless eigenstates.

The free-streaming of massive neutrinos causes suppression of $P^{\rm L}(k)$ relative to the case with massless neutrinos for scales smaller than the neutrino free-streaming scale after the time at which massive neutrinos become non-relativistic -- see Fig.~1 of \cite{FVN} for an example. A linear approximation gives the amplitude of suppression to be $-8f_{\nu}$ where $f_{\nu}=\Omega_{\nu}/\Omega_{\rm m}$ is the fraction of total matter in massive neutrinos \cite{MassiveNeutrinoReview}. Scale-dependent suppression also affects $f(k)$, although the amplitude of this effect is much smaller, as can be seen in Fig.~5 of \cite{FVN}.

The expressions for the 1-loop power spectra corrections in terms of the $z=0$ linear power spectrum $P_0(k)=P^{\rm L}(k, z=0)$ were given in Section~\ref{ssec:LCDM_imp}. For our implementation, we want to take $P^{\rm L}(k, z)$ and $f(k, z)$ at the intended \MGcopter{} output redshift from \texttt{CAMB}/\texttt{MGCAMB} and use it as input to \MGcopter{}. Therefore we need to rewrite the expressions for the 1-loop power spectra in terms of $P^{\rm L}(k, z)$ instead of $P_0(k)$, using $F_1(k)=G_1(k)/f(k, z)$ and $P_0(k)=P^{\rm L}(k, z)/F_1^2(k)=f^2(k, z)P^{\rm L}(k, z)/G_1^2(k,z)$. The 22 correction terms are
\\
\begin{align}
    P_{\delta\delta}^{22}(k) =& 2 \frac{k^3}{(2\pi)^2}\int_0^{\infty}r^2 \mathrm{d}r \int_{-1}^1 P^{\rm L}(kr, z)P^{\rm L}(k\sqrt{1+r^2-2rx}, z)
    \nonumber \\ &\times \frac{F_2^2(k, r, x)}{F_1^2(kr)F_1^2(k\sqrt{1+r^2-2rx})} \mathrm{d}x~,
    \\
    \nonumber \\
    P_{\delta\theta}^{22}(k) =& 2 \frac{k^3}{(2\pi)^2} \int_0^{\infty}r^2 \mathrm{d}r \int_{-1}^1 P^{\rm L}(kr, z)P^{\rm L}(k\sqrt{1+r^2-2rx}, z)
    \nonumber \\ &\times f(kr, z) f(k\sqrt{1+r^2-2rx}, z) \frac{G_2(k, r, x)}{G_1(kr) G_1(k\sqrt{1+r^2-2rx})}
    \nonumber \\ &\times \frac{F_2(k, r, x)}{F_1(kr)F_1(k\sqrt{1+r^2-2rx})}\mathrm{d}x~,
    \\
    \nonumber \\
    P_{\theta\theta}^{22}(k) =& 2 \frac{k^3}{(2\pi)^2} \int_0^{\infty}r^2 \mathrm{d}r \int_{-1}^1 P^{\rm L}(kr, z)P^{\rm L}(k\sqrt{1+r^2-2rx}, z)
    \nonumber \\ &\times f^2(kr, z)f^2(k\sqrt{1+r^2-2rx}, z) \frac{G_2^2(k, r, x)}{G_1^2(kr) G_1^2(k\sqrt{1+r^2-2rx})} \mathrm{d}x~,
\end{align}
\\
while the 13 correction terms are
\\
\begin{align}
    P_{\delta\delta}^{13}(k) =& 6 \frac{k^3}{(2\pi)^2} P^{\rm L}(k, z) \int_0^{\infty} r^2 P^{\rm L}(kr, z) \frac{F_3(k, r, x)}{F_1(k)F_1^2(kr)} \mathrm{d}r~,
    \\
    \nonumber \\
    P_{\delta\theta}^{13}(k) =& 3 \frac{k^3}{(2\pi)^2}F_1(k) P^{\rm L}(k, z) \int_0^{\infty} r^2 P^{\rm L}(kr, z) f(k, z)f^2(kr, z) \frac{G_3(k, r, x)}{G_1(k)G_1^2(kr)} \mathrm{d}r
    \nonumber \\ &+ 3 \frac{k^3}{(2\pi)^2}f(k, z)P^{\rm L}(k, z) \int_0^{\infty} r^2 P^{\rm L}(kr, z) \frac{F_3(k, r, x)}{F_1(k)F_1^2(kr)} \mathrm{d}r~,
    \\
    \nonumber \\
    P_{\theta\theta}^{13} =& 6 \frac{k^3}{(2\pi)^2}P^{\rm L}(k, z) \int_0^{\infty} r^2 P^{\rm L}(kr, z) f^2(k, z)f^2(kr, z) \frac{G_3(k, r, x)}{G_1(k)G_1^2(kr)} \mathrm{d}r~.
\end{align}
\\
Note that in these expressions the only terms to contain massive neutrinos are $P^{\rm L}$ and $f$; all of the kernels $F_i$ and $G_i$ are unmodified. The $A$ and $B$ terms written in \cref{eq:A,eq:B} are also computed as convolutions of two linear power spectra with kernels, and thus are rewritten using the same method as for $P^{13}$ and $P^{22}$. We have implemented these equations in \MGcopter{}.

\section{Validation} \label{sec:Validation}

In order to validate our implementation of massive neutrinos in the \MGcopter{} code, we have tested its output against results from the fast, approximate {\it N}-body code \MGpicola{}, which is a modified version of \picola{} \cite{LPICOLA} that includes modified gravity \cite{Winther2017} and massive neutrinos \cite{Wrightetal2017} and has been tested against full {\it N}-body simulations. In the legends of the figures that follow we shall refer to our modified \MGcopter{} code simply as \copter{}, and the \MGpicola{} code as \cola{}.

Throughout, we use paired-fixed \MGpicola{} simulations where we produce two simulations with fixed amplitudes, meaning the initial amplitudes of the Fourier modes of the density field are set to that of the ensemble average power spectrum, and paired, where the initial modes in the second simulation are mirrored compared to those of the first \cite{PairedFixed}. This procedure significantly reduces variance that arises from the sparse sampling of wavemodes without the need for averaging over a large number of density field realisations, and has been shown not to introduce a bias to the recovery of the mean properties of the Gaussian ensemble, despite the fixing introducing non-Gaussianity \cite{PairedFixedAccuracy}. However, we also ran five additional \MGpicola{} simulations for each model with randomised realisations of the initial density field. The standard deviation in the power spectra of these additional five simulations is used for the error bars in the figures below unless explicitly stated otherwise. The modified gravity model considered here is the Hu-Sawicki $f(R)$ model, which has one free parameter \frz{} and we refer to $|f_{R0}|=10^{-5}$ and $|f_{R0}|=10^{-4}$ as F5 and F4 respectively. The velocity divergence field $\theta$ has been computed using the \texttt{DTFE} code \cite{DTFE}. The cosmological parameters used in this paper are the same as in \cite{Baldi2014}; $h=0.671$, $\Omega_{\rm m}=0.3175$, $\Omega_{\rm b}=0.049$, $A_s=2.215\times 10^{-9}$, and $n_{\rm s}=0.966$.

Note that a recent version update of \texttt{MGCAMB} improved the handling of massive neutrinos \cite{MGCAMB3}. Although our results were produced using the previous version of \texttt{MGCAMB}, we have verified that for the parameters we use the difference in the linear power spectrum between the two versions is negligible.

\begin{figure*}[t]
\begin{center}
\includegraphics[width=\textwidth]{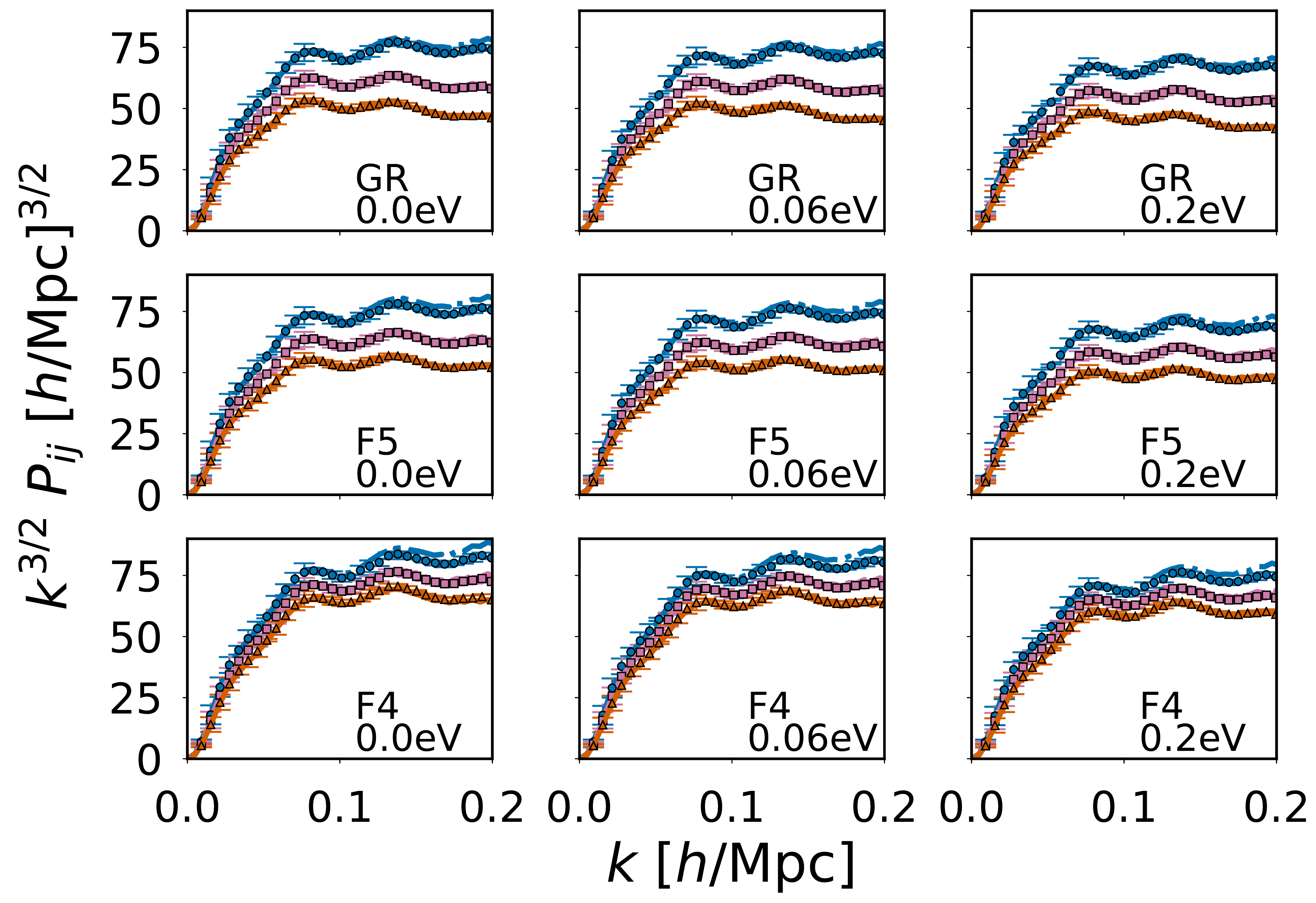}
\caption[]{\small Real-space non-linear power spectra for various gravity models and neutrino masses at $z=1$. Points represent the results of paired-fixed \MGpicola{} {\it N}-body simulations, while lines are the result of \MGcopter{}. The blue circles and dashed-dotted line give the density auto-correlated power spectra $P_{\delta\delta}$, the pink squares and dashed line give the density-velocity divergence cross-correlated power spectra $P_{\delta\theta}$, while the orange triangles and solid line give the velocity divergence auto-correlated power spectra $P_{\theta\theta}$.}
\label{fig:real}
\vspace{-3ex}
\end{center}
\end{figure*}

\begin{figure*}[t]
\begin{center}
\includegraphics[width=\textwidth]{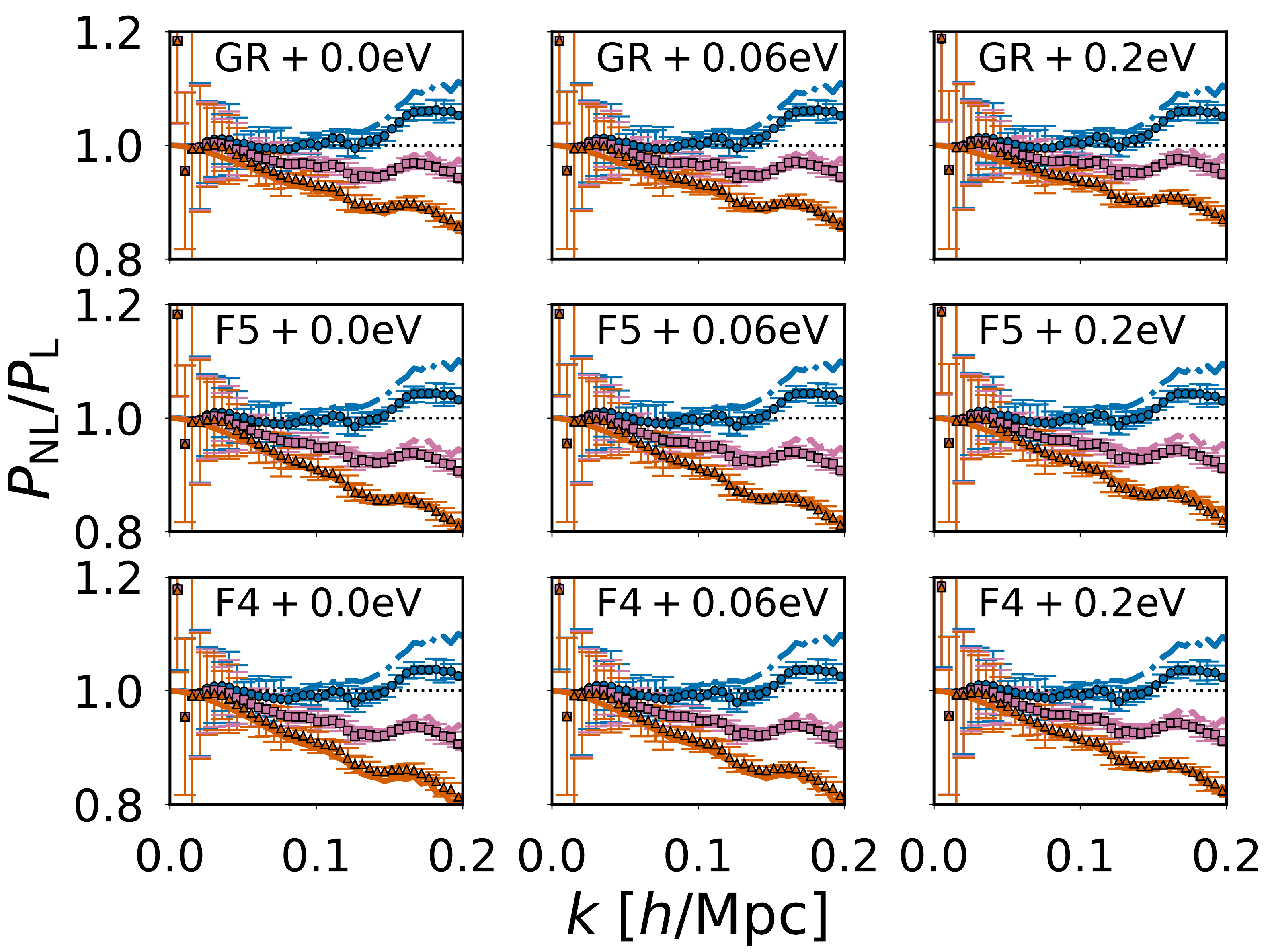}
\caption[]{\small As in Fig.~\ref{fig:real} but for the ratio of the real-space non-linear power spectra to their linear counterparts.}
\label{fig:real_NLL}
\vspace{-3ex}
\end{center}
\end{figure*}

\begin{figure*}[t]
\begin{center}
\includegraphics[width=\textwidth]{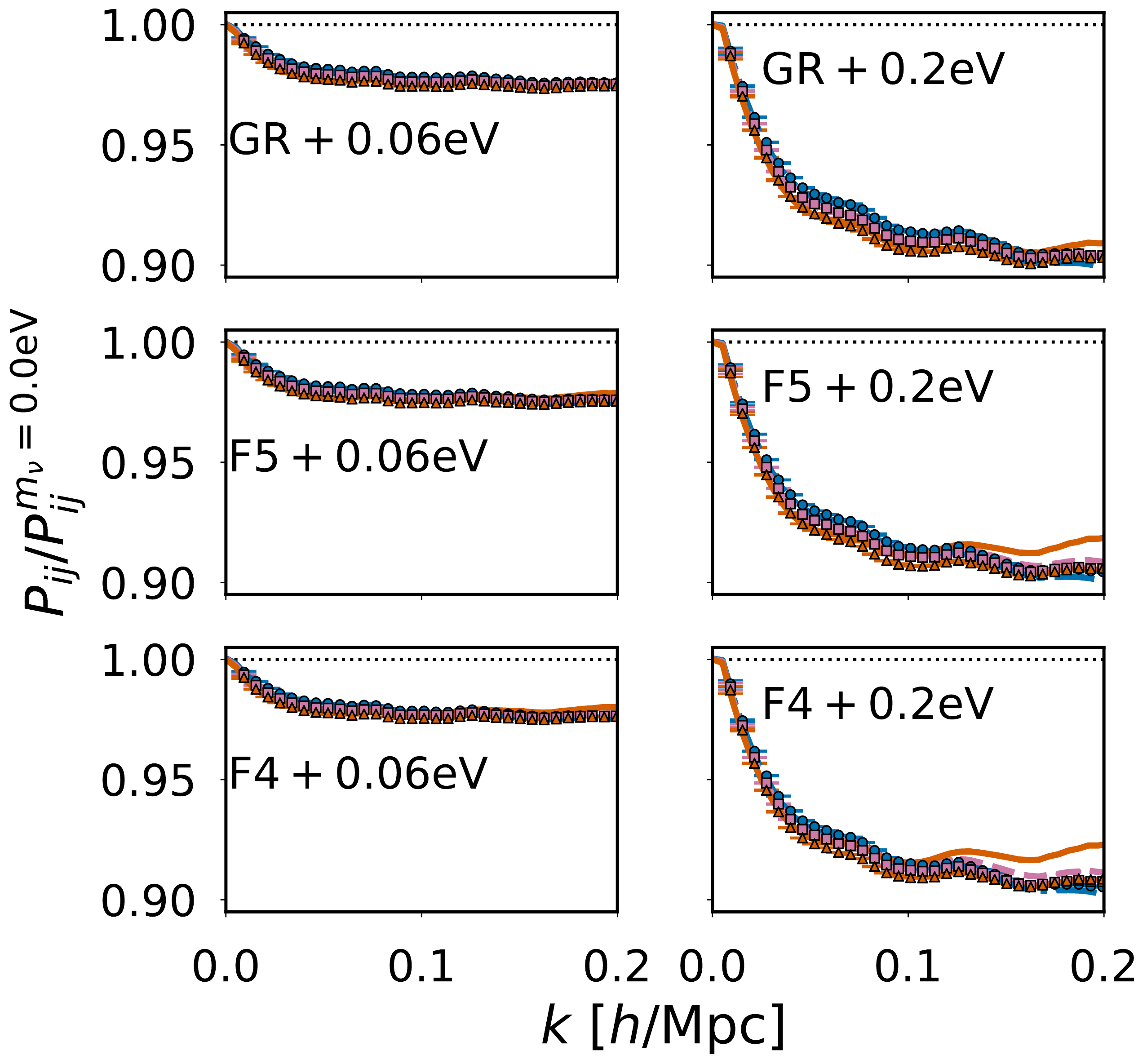}
\caption[]{\small As in Fig.~\ref{fig:real}, but for the ratio of real-space non-linear power spectra with and without neutrino mass.}
\label{fig:real_ratio}
\vspace{-3ex}
\end{center}
\end{figure*}

We first study the comparison between \MGcopter{} and \MGpicola{} in the real-space power spectra, in \cref{fig:real,fig:real_NLL,fig:real_ratio}. Figure~\ref{fig:real} shows the real-space non-linear power spectra at $z=1$ computed with both \MGpicola{} and \MGcopter{}. We display the density auto-correlation $P_{\delta\delta}$, the velocity divergence auto-correlation $P_{\theta\theta}$, and the density-velocity divergence cross-correlation $P_{\delta\theta}$, in the form $k^{3/2}P_{ij}$ for ease of viewing, for GR, F5, and F4 each with $0.0$eV, $0.06$eV, and $0.2$eV neutrinos. The error bars on the (paired-fixed) \MGpicola{} points are the standard deviation of the 5 additional (non-paired-fixed) \MGpicola{} simulations. In all cases, \MGcopter{} reproduces the results of the \MGpicola{} simulations very well up to the start of the quasi-non-linear scale around $k=0.1~h/{\rm Mpc}$ where perturbation theory begins to break down. The agreement between \MGcopter{} and \MGpicola{} persists to larger $k$ values for $P_{\theta\theta}$ and $P_{\delta\theta}$ than $P_{\delta\delta}$, which is consistent with the behaviour seen when \MGcopter{} was compared to full {\it N}-body simulations in Fig.~10 of \cite{BoseKoyama2016}.

Figure~\ref{fig:real_NLL} displays the same data but presented as the ratio of the full non-linear power spectra to their linear components, which helps to show where the modelling of non-linearities with \MGcopter{} becomes inaccurate. Figure~\ref{fig:real_ratio} again shows the same data but presented as the ratio of the power-spectra with and without massive neutrinos for both the $0.06$eV and $0.2$eV neutrinos. The scale up to which \MGcopter{} closely follows the results of the \MGpicola{} simulations is marginally improved due to taking the ratio between power spectra in two models.

\begin{figure*}[t]
\begin{center}
\includegraphics[width=\textwidth]{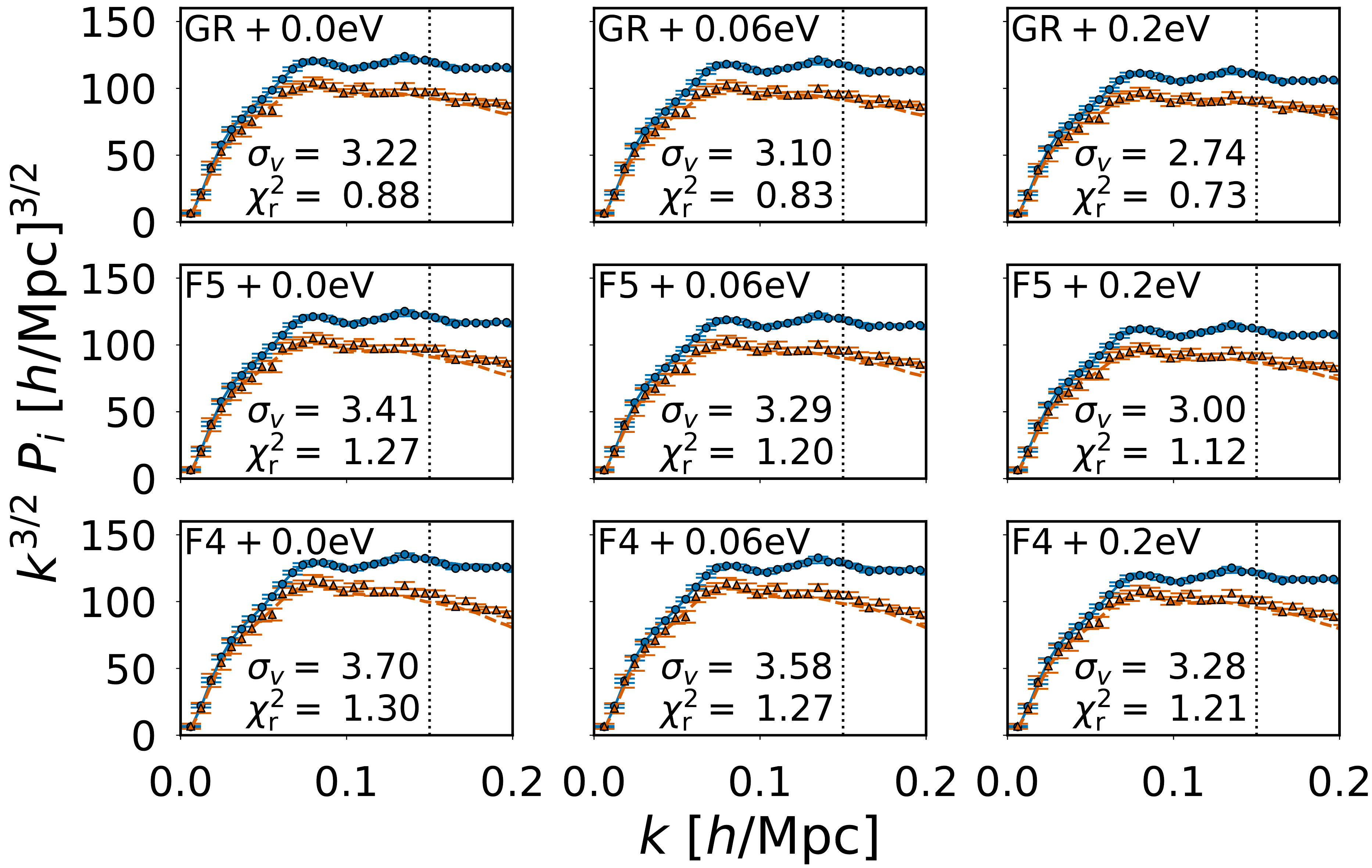}
\caption[]{\small Redshift-space non-linear power spectra for various gravity models and neutrino masses at $z=1$. Points represent the results of paired-fixed \MGpicola{} {\it N}-body simulations, while solid lines are the result of \MGcopter{} with velocity dispersion $\sigma_{v}$ fitted to \MGpicola{} up to $k=0.15~h/{\rm Mpc}$, shown by the vertical dashed line. The error bars are those of an ideal survey with survey volume $V_{\rm s} = 10~{\rm Gpc}^3/h^3$ and galaxy density $\bar{n}_{\rm g}=4\times 10^{-3}~h^3/{\rm Mpc}^3$. The blue circles and solid line give the monopole $P_0$, and the orange squares and dashed line give the quadrupole $P_2$.}
\label{fig:red}
\vspace{-3ex}
\end{center}
\end{figure*}

\begin{figure*}[t]
\begin{center}
\includegraphics[width=\textwidth]{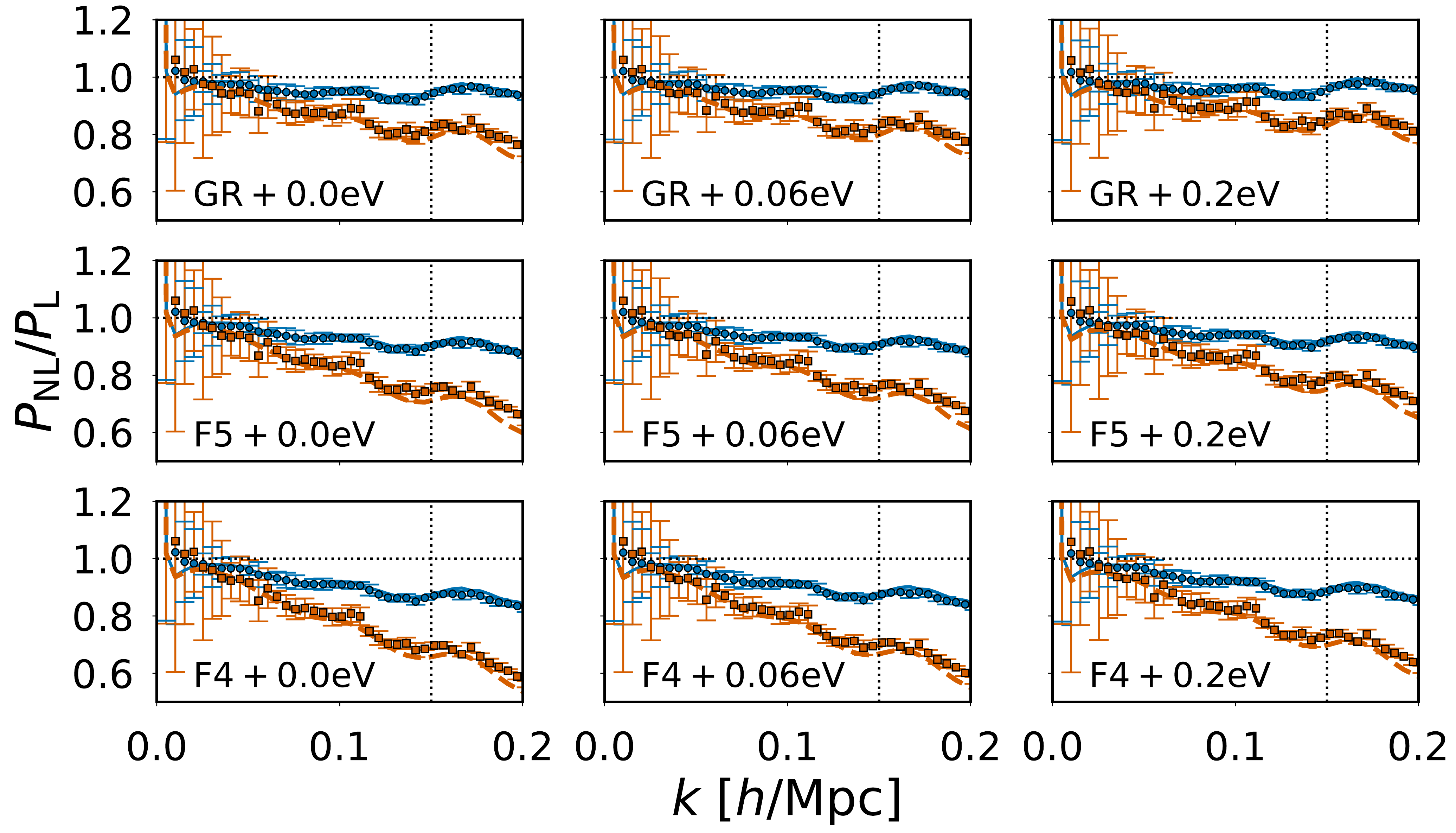}
\caption[]{\small As in Fig.~\ref{fig:red} but for the ratio of the redshift-space power spectrum multipoles to their linear (Kaiser) counterparts. The error bars on the \MGpicola{} points represent the standard deviation of the five additional \MGpicola{} simulations.}
\label{fig:red_NLL}
\vspace{-3ex}
\end{center}
\end{figure*}

\begin{figure*}[t]
\begin{center}
\includegraphics[width=\textwidth]{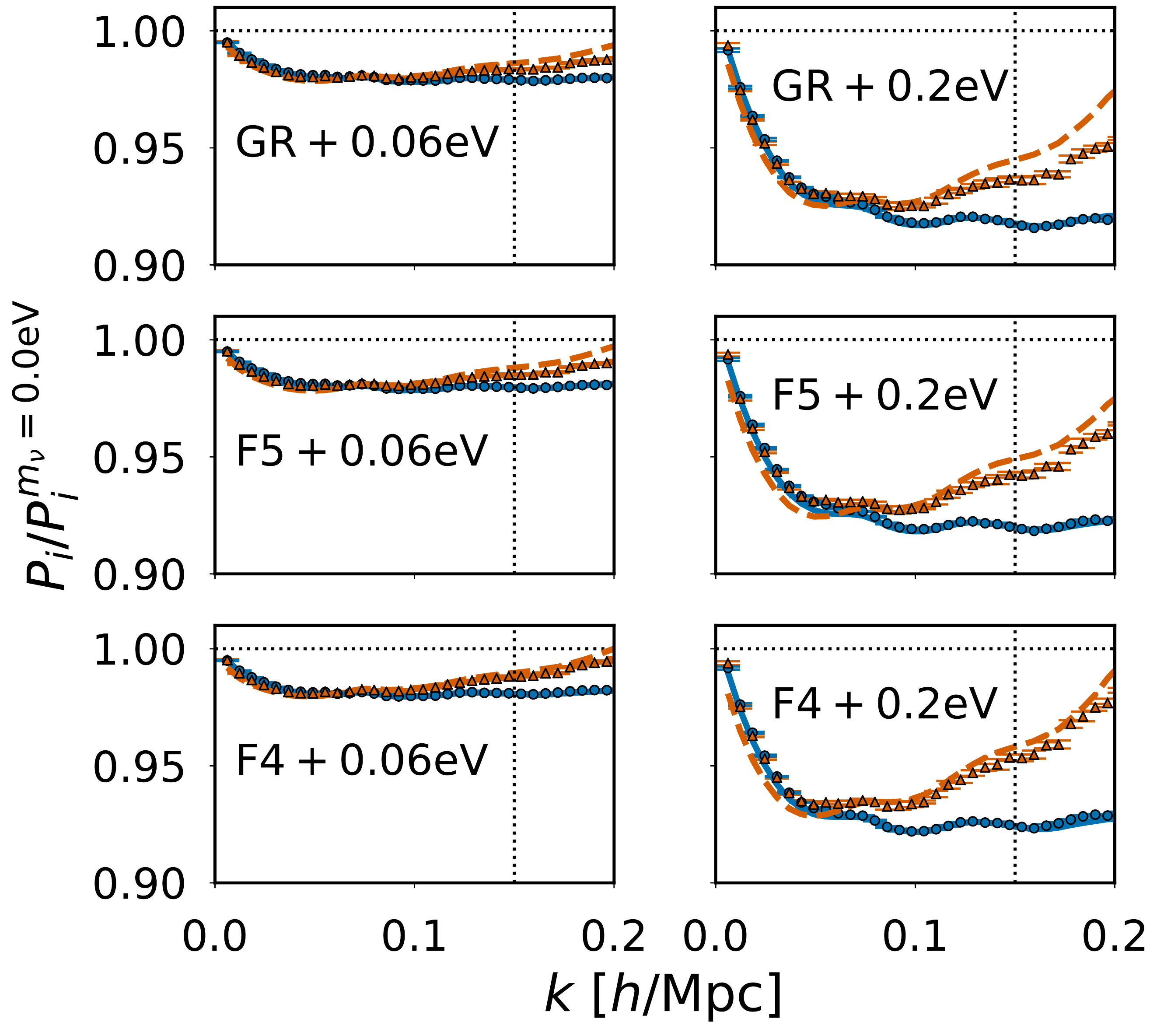}
\caption[]{\small As in Fig.~\ref{fig:red}, but for the ratio of redshift-space non-linear power spectra with and without neutrino mass. The error bars on the \MGpicola{} points represent the standard deviation of the five additional \MGpicola{} simulations.}
\label{fig:red_ratio}
\vspace{-3ex}
\end{center}
\end{figure*}

Next, we look at the comparison between \MGpicola{} and \MGcopter{} with $\sigma_{v}$ fitted to the \MGpicola{} simulations in the non-linear redshift-space power spectra in \cref{fig:red,fig:red_NLL,fig:red_ratio}. Figure~\ref{fig:red} shows the monopole $P_0$ and quadrupole $P_2$ of the redshift-space power spectra for GR, F5, and F4 gravity models each with $0.0$eV, $0.06$eV, and $0.2$eV neutrinos. We display the results computed from paired-fixed \MGpicola{} simulations and \MGcopter{} with the TNS velocity dispersion parameter $\sigma_v$ fitted to the \MGpicola{} simulations up to $k=0.15~h/{\rm Mpc}$ in the form $k^{3/2}P_i(k)$; the figure includes the best-fitting values of $\sigma_v$ (expressed in RSD displacement units ${\rm Mpc}/h$) and the reduced $\chi^2$ for each model. The error bars on the \MGpicola{} points are taken from the inverse covariance matrices used in the $\sigma_v$ fitting procedure, whose computation is described at the end of Section~\ref{ssec:LCDM_imp}. The $\sigma_v$ fitting procedure prioritises recovering the monopole $P_0$, and thus the agreement between \MGcopter{} and \MGpicola{} is slightly worse for the quadrupole $P_2$. As expected, for each gravity model increasing the mass of the neutrinos leads to a decrease in the best-fitting value of $\sigma_v$ and the quality of the fit increases, while for a fixed neutrino mass increasing the strength of the modification of gravity from GR to F5 and then F4 leads to an increase in the best-fitting value of $\sigma_v$ and a slightly worse quality of fit. The reason for this behaviour is that enhancement to gravity leads to an increase in the velocities of galaxies around an overdensity, thus increasing the non-linear damping, while massive neutrinos have the opposite effect due to their suppression of structure formation. The quality of the fit is better when the non-linearity is smaller and vice versa. However, in all cases the quality of the fit of \MGcopter{} to \MGpicola{} is good up to quasi-non-linear scales.

Figure~\ref{fig:red_NLL} displays the same data as Fig.~\ref{fig:red} but presented as the ratio of the full non-linear multipoles to their linear counterparts computed with the Kaiser RSD model \cite{Kaiser1987}, while Fig.~\ref{fig:red_ratio} presents the data of Fig.~\ref{fig:red} as the ratio of the non-linear power-spectra multipoles with and without massive neutrinos for both the $0.06$eV and $0.2$eV neutrinos. The error bars on the \MGpicola{} points in these two figures represent the standard deviation of the 5 additional \MGpicola{} simulations. As in real-space, the scale up to which \MGcopter{} closely follows the results of the \MGpicola{} simulations is slightly improved due to taking the ratio between power spectra in two models.

\begin{figure*}[t]
\begin{center}
\includegraphics[width=\textwidth]{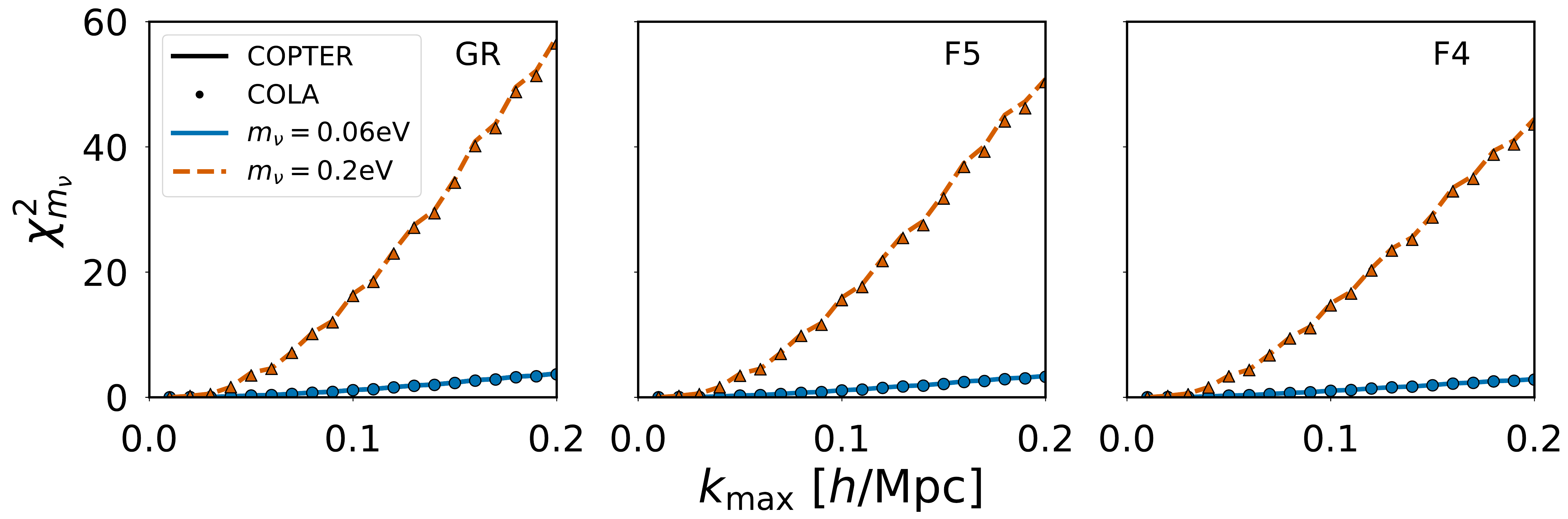}
\caption[]{\small Difference between the redshift-space multipoles with and without neutrino mass $\chi^2_{m_{\nu}}$ as a function of the maximum scale $k_{\rm max}$ at $z=1$ for GR in the left panel, F5 in the middle panel, and F4 in the right panel. Points represent the results of paired-fixed \MGpicola{} {\it N}-body simulations, while solid lines are the result of \MGcopter{} with velocity dispersion $\sigma_{v}$ fitted to \MGpicola{} up to $k=k_{\rm max}$. Blue data corresponds to $m_{\nu}=0.06\mathrm{eV}$, and orange to $m_{\nu}=0.2\mathrm{eV}$.}
\label{fig:chi_mnu}
\vspace{-3ex}
\end{center}
\end{figure*}

We also quantify the ability of \MGcopter{} to recover the redshift-space multipole results of \MGpicola{} through $\chi^2_{m_{\nu}}$; the difference between the redshift-space multipoles with and without neutrino mass. In Fig.~\ref{fig:chi_mnu} we display $\chi^2_{m_{\nu}}$ as a function of the maximum comparison scale $k_{\rm max}$ for GR, F5, and F4 each with $0.06$eV and $0.2$eV neutrinos at $z=1$. Here, \MGcopter{} is fitted to the \MGpicola{} simulations up to $k_{\rm max}$ with the covariance computed assuming an ideal survey as described at the end of Section~\ref{ssec:LCDM_imp}. The agreement in $\chi^2_{m_{\nu}}$ between \MGpicola{} and \MGcopter{} fitted to \MGpicola{} is excellent in all cases. This implies that \MGcopter{} with $\sigma_v$ fitted to simulations is capable of capturing the effect of massive neutrinos accurately.

\section{Degeneracy} \label{sec:Degeneracy}

With the inclusion of modified gravity and massive neutrinos in \MGcopter{} the degeneracy between the two effects can be investigated.

\subsection{Real- and redshift-space}

We start by studying the degeneracy between modified gravity and massive neutrinos in real space.

\begin{figure*}[t]
\begin{center}
\includegraphics[width=\textwidth]{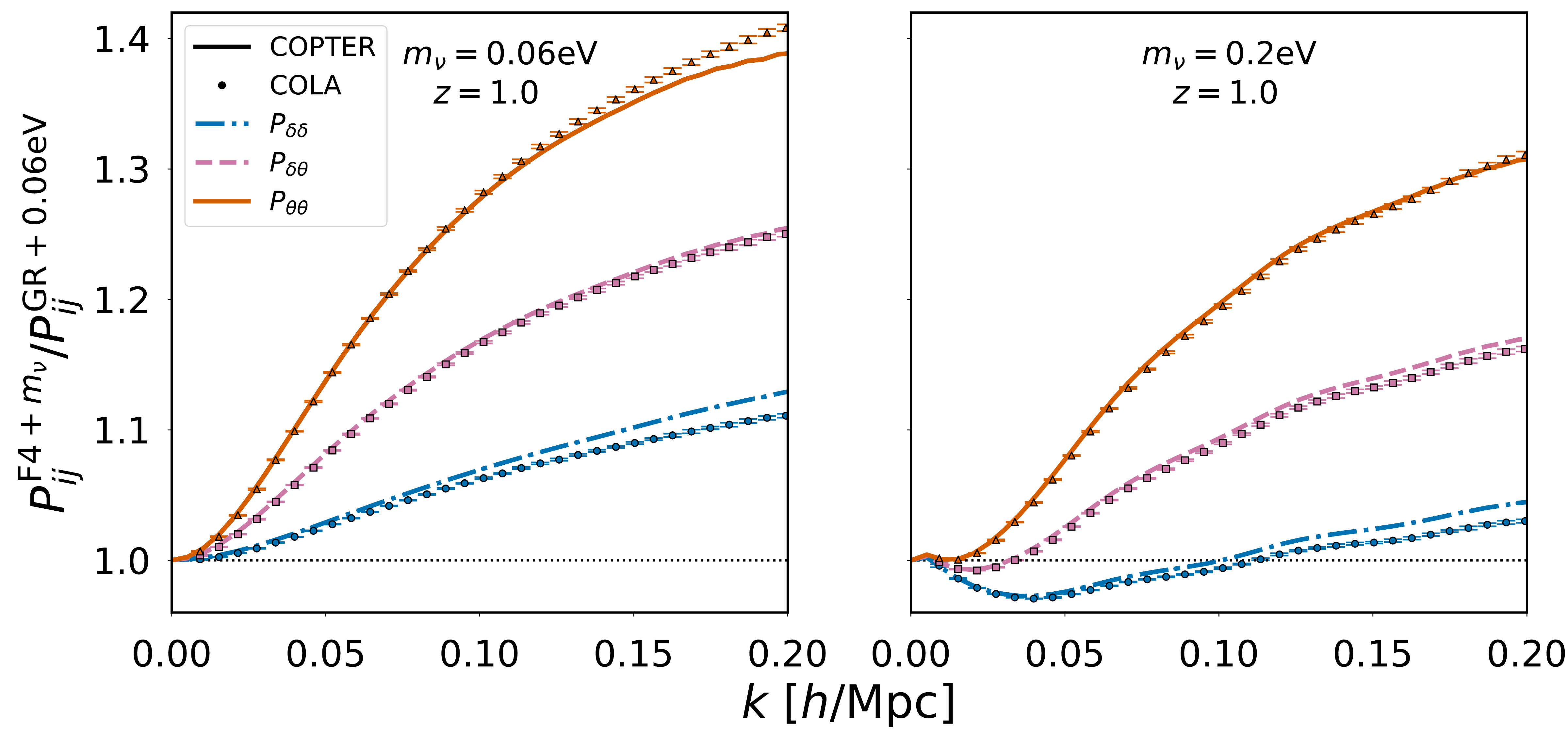}
\caption[]{\small Ratio of real-space power spectra in F4 with $m_{\nu}=0.06$eV (left panel) and $m_{\nu}=0.2$eV (right panel) to the fiducial model of GR with $m_{\nu}=0.06$eV at $z=1$. Points represent the results of paired-fixed \MGpicola{} {\it N}-body simulations, while lines are the result of \MGcopter{}. The blue circles and dashed-dotted line give the density auto-correlated power spectra $P_{\delta\delta}$, the pink squares and dashed line give the density-velocity divergence cross-correlated power spectra $P_{\delta\theta}$, while the orange triangles and solid line give the velocity divergence auto-correlated power spectra $P_{\theta\theta}$.}
\label{fig:degen_real_comp}
\vspace{-3ex}
\end{center}
\end{figure*}

In Fig.~\ref{fig:degen_real_comp} we display the ratio of real-space power spectra in F4 gravity with $0.06$eV neutrinos in the left panel and $0.2$eV neutrinos in the right panel to a fiducial model which we take to be GR with $0.06$eV neutrinos at $z=1$. We show results for the density auto-correlation $P_{\delta\delta}$, the velocity divergence auto-correlation $P_{\theta\theta}$, and the density-velocity divergence cross-correlation $P_{\delta\theta}$. The results of paired-fixed \MGpicola{} simulations and of \MGcopter{} are plotted. The error bars on the \MGpicola{} results are computed using the standard deviation over the five additional simulations. In all cases, the results of \MGcopter{} agree well with those of \MGpicola{} up to quasi-non-linear scales around $k=0.1~h/{\rm Mpc}$. The left panel, where the neutrino masses are the same in both GR and F4, shows the scale-dependent enhancement of the real-space power spectra provided by F4 gravity. However, when heavier neutrinos are added to the F4 case, as in the right panel, this enhancement is opposed by the suppression effect of the massive neutrinos. Indeed, the right panel shows that $P_{\delta\delta}$ is a poor probe to distinguish between GR with $0.06$eV neutrinos and F4 with $0.2$eV neutrinos in this particular case. However, the two models remain distinguishable in $P_{\delta\theta}$ and $P_{\theta\theta}$, showing that velocity information has the potential to break the degeneracy between modified gravity and massive neutrinos. This was recently shown using the results of full {\it N}-body simulations \cite{Hagstotzetal2019}. However, neither $P_{\delta\theta}$ nor $P_{\theta\theta}$ can be measured directly by observations. Instead, it is necessary to extract the velocity information that is encoded within redshift-space distortions, and it is to this we turn our attention. We shall refer to GR with $0.06$eV neutrinos and F4 with $0.2$eV neutrinos as our two degenerate models.

\begin{figure*}[t]
\begin{center}
\includegraphics[width=\textwidth]{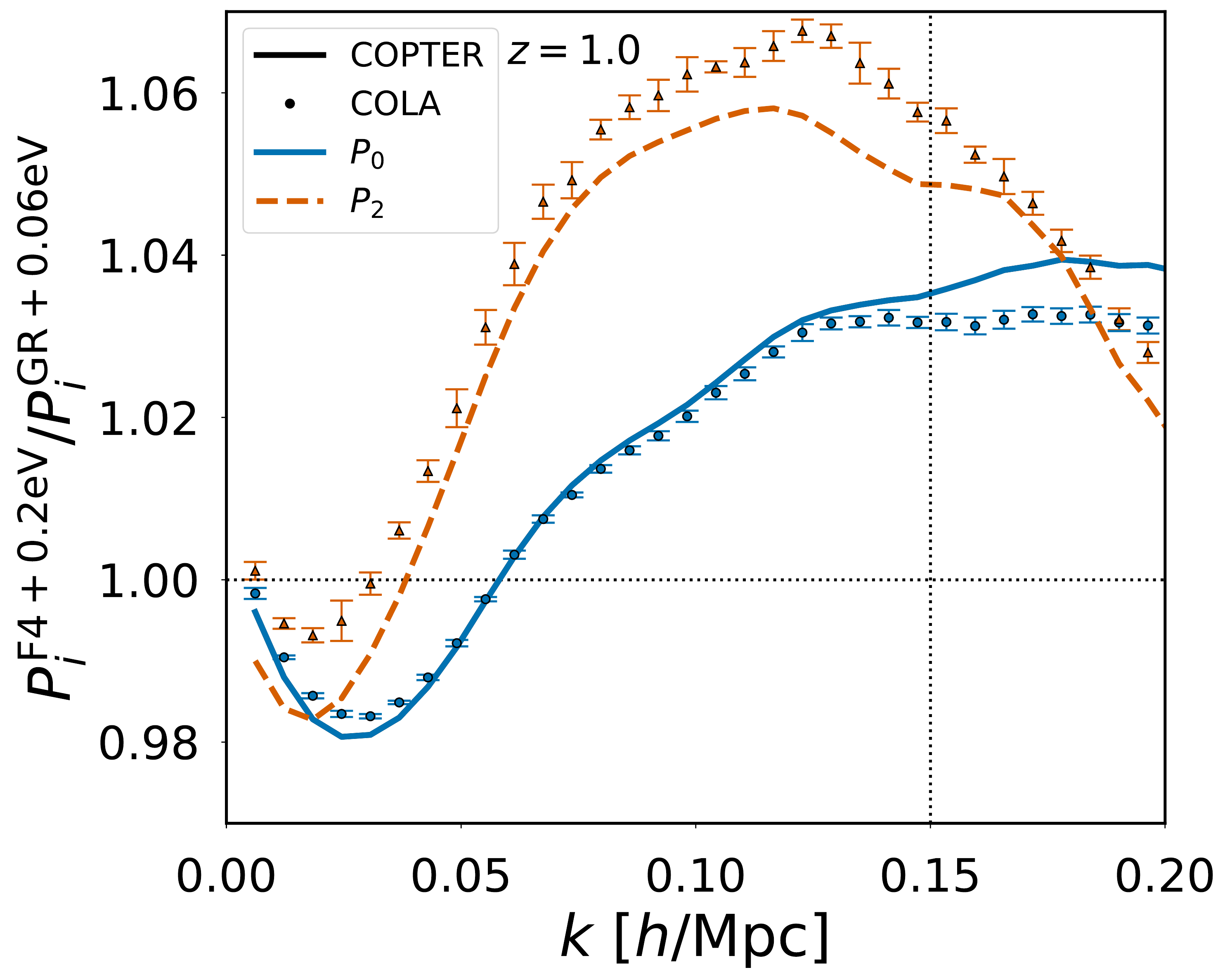}
\caption[]{\small Degeneracy between F4 with $m_{\nu}=0.2$eV and the fiducial model of GR with $m_{\nu}=0.06$eV in the redshift-space power spectrum multipoles at $z=1$, represented as the ratio of power spectra in the two models. Points represent the results of paired-fixed \MGpicola{} {\it N}-body simulations, while solid lines are the result of \MGcopter{} with velocity dispersion $\sigma_{v}$ fitted to \MGpicola{} up to $k=0.15~h/{\rm Mpc}$. The blue circles and solid line give the monopole $P_{0}$, while the orange squares and dashed line give the quadrupole $P_{2}$. The best-fitting value of $\sigma_v$ for each model and the associated reduced $\chi^2$ are $\sigma_v=3.07~{\rm Mpc}/h$ with $\chi^2_{\rm r}=0.27$ for GR+$0.06$eV and $\sigma_v=3.23~{\rm Mpc}/h$ with $\chi^2_{\rm r}=0.39$ for F4+$0.2$eV.}
\label{fig:degen_red}
\vspace{-3ex}
\end{center}
\end{figure*}

In Fig.~\ref{fig:degen_red} we plot the redshift-space monopole and quadrupoles in F4 gravity with $0.2$eV neutrinos normalised to GR with $0.06$eV neutrinos computed with both \MGcopter{} and \MGpicola{}. For each model the \MGcopter{} result has been produced by fitting $\sigma_v$ to the paired-fixed \MGpicola{} simulation up to $k=0.15~h/{\rm Mpc}$ with the covariance computed assuming a DESI-like survey as detailed at the end of Section~\ref{ssec:LCDM_imp}. The error bars on the \MGpicola{} results are computed using the standard deviation over five simulations with a boxsize of $1~{\rm Gpc}/h$ for each model. Firstly, this plot shows that modelling the redshift-space monopole and quadrupole using \MGcopter{} with $\sigma_v$ fitted to \MGpicola{} simulations works well. Secondly, for our degenerate models, while the monopole is still a poor probe for distinguishing between the models, the quadrupole, by virtue of the encoding of velocity information, displays differences between the two models and thus has the potential to break the degeneracy.

\subsection{Redshift evolution}

Our method also allows us to investigate how the degeneracy evolves with redshift in both real- and redshift-space.

\begin{figure*}[t]
\begin{center}
\includegraphics[width=\textwidth]{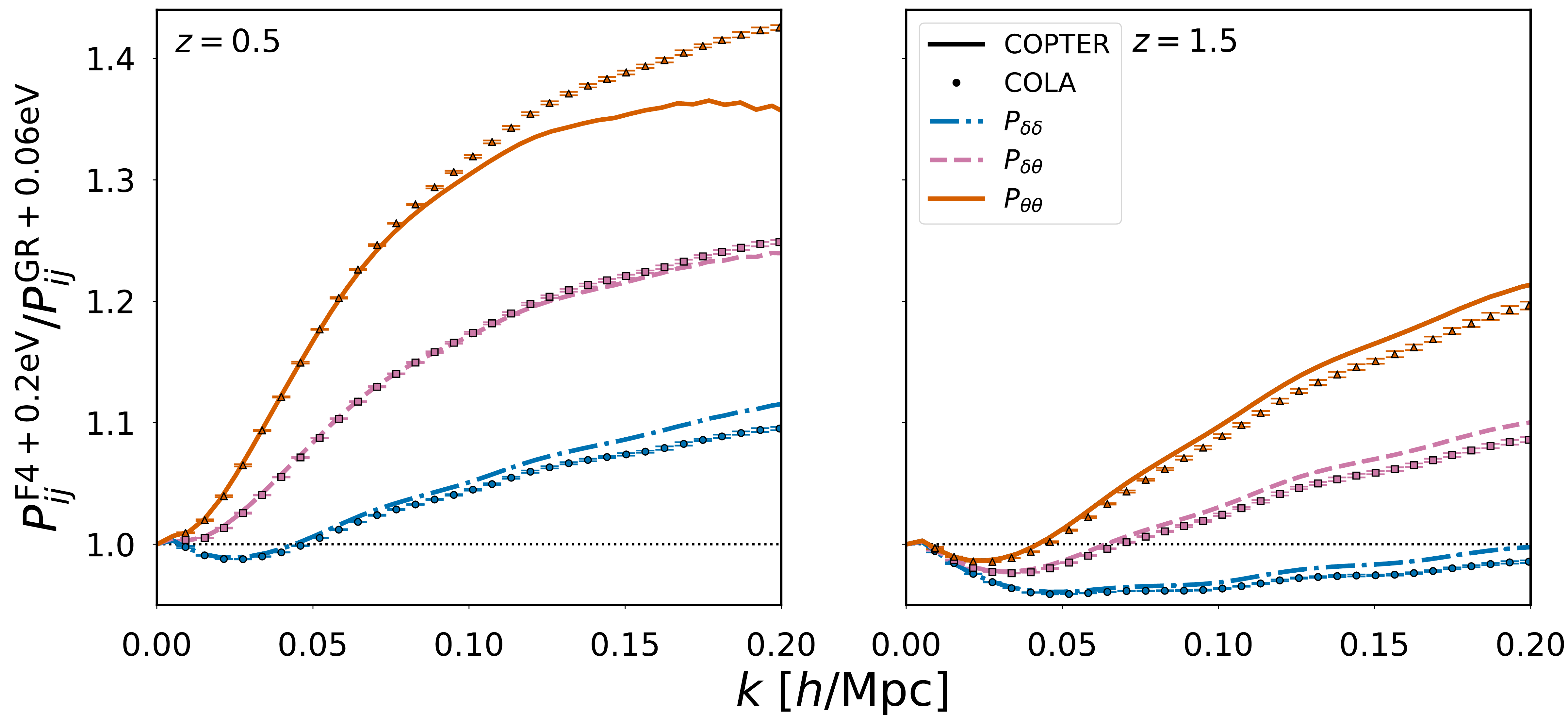}
\caption[]{\small As in the right panel of Fig.~\ref{fig:degen_real_comp}, but showing the evolution of the degeneracy with redshift. The left panel corresponds to $z=0.5$ and the right to $z=1.5$.}
\label{fig:degen_real_zevo}
\vspace{-3ex}
\end{center}
\end{figure*}

\begin{figure*}[t]
\begin{center}
\includegraphics[width=\textwidth]{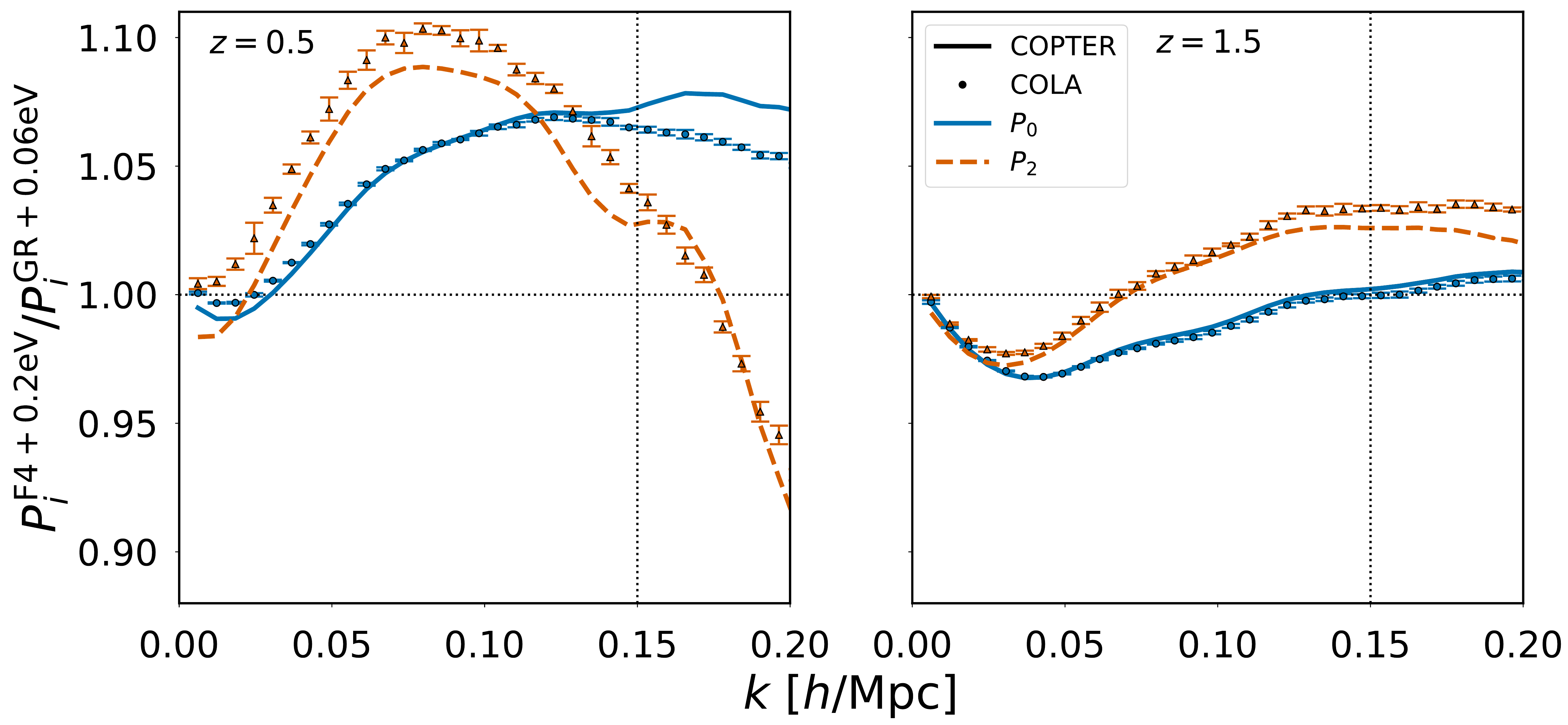}
\caption[]{\small As in Fig.~\ref{fig:degen_red}, but showing the evolution of the degeneracy with redshift. The left panel corresponds to $z=0.5$ and the right to $z=1.5$. For GR with $0.06$eV neutrinos, the best-fitting value of $\sigma_v$ and the corresponding reduced $\chi^2$ are $\sigma_v=3.84~{\rm Mpc}/h$ and $\chi^2_{\rm r}=0.29$ for $z=0.5$, and $\sigma_v=2.36~{\rm Mpc}/h$ and $\chi^2_{\rm r}=0.065$ for $z=1.5$. For F4 with $0.2$eV neutrinos, the best-fitting value of $\sigma_v$ and the corresponding reduced $\chi^2$ are $\sigma_v=4.13~{\rm Mpc}/h$ and $\chi^2_{\rm r}=0.34$ for $z=0.5$, and $\sigma_v=2.45~{\rm Mpc}/h$ and $\chi^2_{\rm r}=0.079$ for $z=1.5$. In all cases $\sigma_v$ has been fitted to \MGpicola{} up to $k=0.15~h/{\rm Mpc}$.}
\label{fig:degen_red_zevo}
\vspace{-3ex}
\end{center}
\end{figure*}

In Fig.~\ref{fig:degen_real_zevo} we show the real-space power spectra in the ratio between the two degenerate models as in the right panel of Fig.~\ref{fig:degen_real_comp} but at $z=0.5$ (left panel) and $z=1.5$ (right panel). In Fig.~\ref{fig:degen_red_zevo} we show the redshift-space power spectrum multipoles in the ratio between the two degenerate models as in Fig.~\ref{fig:degen_red} but at $z=0.5$ (left panel) and $z=1.5$ (right panel). These figures demonstrate that the degeneracy evolves significantly with redshift, both in real- and redshift-space. Figure~\ref{fig:degen_real_zevo} shows that while our two degenerate models had similar matter power spectra at $z=1$ it is easier to distinguish between the two models with the matter power spectrum at other redshifts.

\begin{figure*}[t]
\begin{center}
\includegraphics[width=\textwidth]{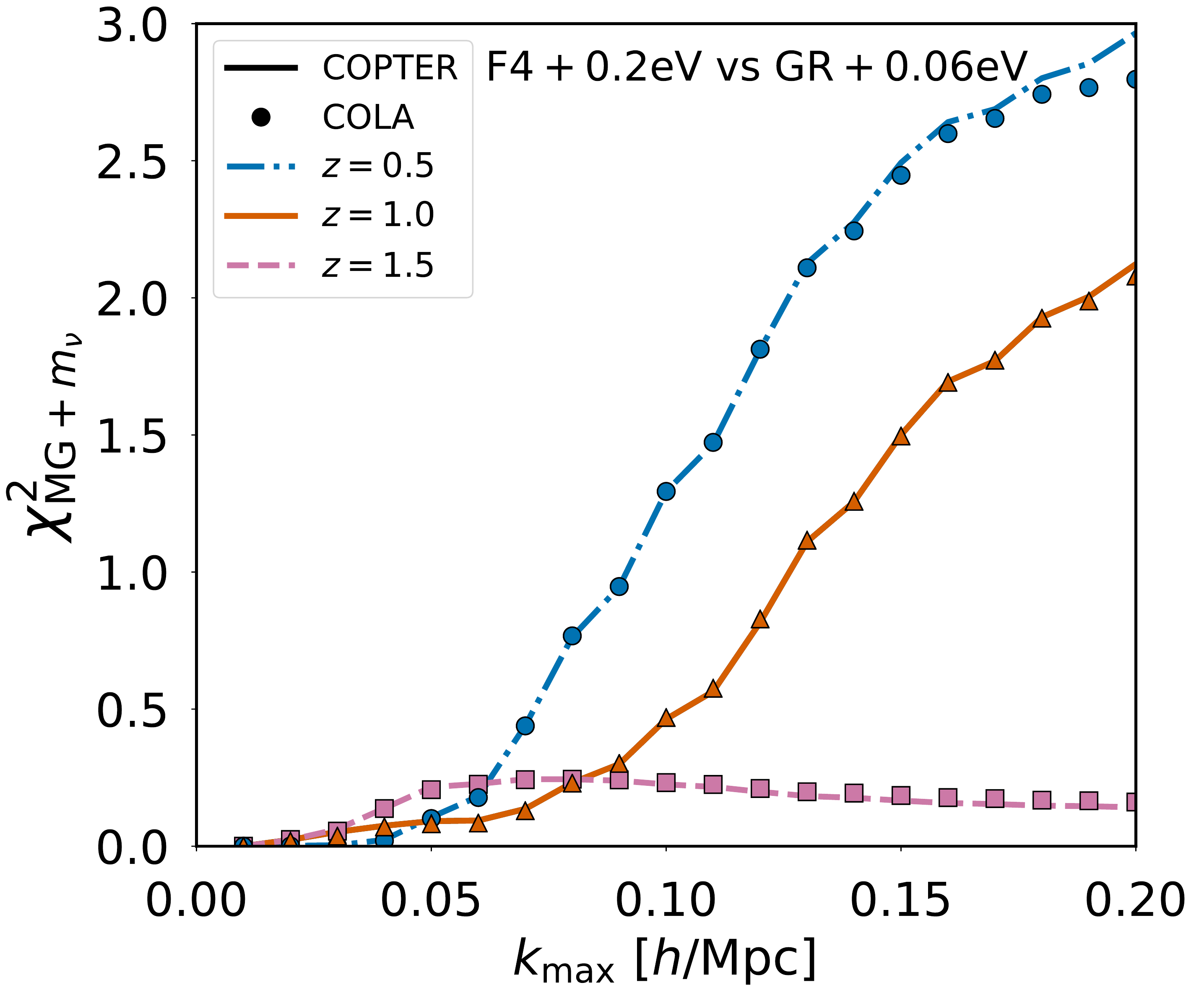}
\caption[]{\small The redshift evolution of $\chi^2_{\mathrm{MG}+m_{\nu}}(k_{\rm max})$ which quantifies the difference between the redshift-space multipoles of the two degenerate models as a function of maximum comparison scale. The blue circles and dashed-dotted line correspond to $z=0.5$, the orange triangles and solid line to $z=1$, and the pink squares and dashed line to $z=1.5$. Points represent the results of paired-fixed \MGpicola{} {\it N}-body simulations, while lines are the result of \MGcopter{} with velocity dispersion $\sigma_{v}$ fitted to \MGpicola{} up to $k=k_{\rm max}$.}
\label{fig:chi_MGmnu_zevo}
\vspace{-3ex}
\end{center}
\end{figure*}

In Fig.~\ref{fig:chi_MGmnu_zevo} we plot the difference between the redshift-space multipoles in the two degenerate models quantified through $\chi^2_{\mathrm{MG}+m_{\nu}}$ as a function of the maximum comparison scale $k_{\rm max}$. We show $\chi^2_{\mathrm{MG}+m_{\nu}}$ as computed by both \MGpicola{} and \MGcopter{} with $\sigma_v$ fitted to \MGpicola{} up to $k_{\rm max}$ with the covariance computed assuming a DESI-like survey as detailed at the end of Section~\ref{ssec:LCDM_imp}. The results from both methods agree with each other very well. We plot $\chi^2_{\mathrm{MG}+m_{\nu}}$ at three redshifts $z=1.5,\ 1.0,\ 0.5$ and it is clear from these results, along with those in \cref{fig:degen_real_zevo,fig:degen_red_zevo}, that the ability to distinguish between the redshift-space multipoles of these two models evolves with redshift. This emphasises the potential for data at multiple redshifts to break the degeneracy. The tomographic nature of weak lensing observations make them well suited to this task, and the combination of redshift-space distortion measurements with weak lensing observations could prove one of the best probes for breaking the modified gravity-massive neutrino degeneracy. However, it should be noted that systematics associated with weak lensing such as baryonic effects and intrinsic alignments may impact the effectiveness of such a probe.

\section{Conclusions}\label{sec:Conclusion}

In this paper, we have studied the potential for redshift-space distortions to break the degeneracy between the enhancement of structure growth provided by modifications to gravity and suppression of structure growth due to massive neutrinos, at the level of the dark matter field. For combinations of modified gravity parameters and neutrino masses that have similar matter power spectra at a given redshift, the growth rates are different and will remain distinguishable. This degeneracy-breaking growth rate information is encoded via velocities into redshift-space distortions. To carry out this work, we have modelled the effects of both modified gravity and massive neutrinos on real- and redshift-space power spectra with Standard Perturbation Theory through the code \MGcopter{}. We find the implementation of modified gravity and massive neutrinos in \MGcopter{} produces a good agreement for both real- and redshift-space power spectra with the simulation results from the code \MGpicola{} in the case of Hu-Sawicki $f(R)$ gravity.

We have then investigated the degeneracy and shown that the quadrupole of the redshift-space power spectrum retains enough of the velocity information to distinguish between GR with light neutrinos and Hu-Sawicki $f(R)$ with heavy neutrinos. The logical next step is to confirm that we can use the computationally inexpensive modelling of RSD in \MGcopter{} to recover a fiducial combination of $|f_{R0}|$ and $m_{\nu}$ from a simulation. An important open question for this endeavour is whether the process of fitting $\sigma_v$ introduces a new degeneracy, where $\sigma_v$ can dampen the redshift-space multipoles of a model with incorrect $|f_{R0}|$ and $m_{\nu}$ values in a way that makes them difficult to distinguish from those of the fiducial simulation. Future work will focus on extending the modelling of RSD with modified gravity and massive neutrinos to dark matter halos and galaxies in order to bring this method closer to being able to use RSD observations to jointly constrain modified gravity and massive neutrinos. It will be important to study how our conclusions change when we consider biased tracers instead of the underlying dark matter, as the bias parameters may also introduce additional degeneracies.

We have also briefly studied how the degeneracy evolves with redshift. There is a clear evolution of the degeneracy with redshift even for the matter power spectrum; for combinations of modified gravity and neutrino mass parameters that give comparable matter power spectra at one redshift, the matter power spectra at another redshift are in general likely to be distinguishable. The tomographic nature of weak lensing is particularly well suited to investigating this approach to breaking the degeneracy, although weak lensing systematics such as baryonic effects and intrinsic alignments could cause complications. Alternatively, if modified gravity is only a low redshift effect, a constraint on neutrino mass from clustering at higher redshift, for example from HI intensity mapping \cite{HINeutrinos}, would help break the degeneracy.

A work that appeared shortly after this paper studied RSD in halos from simulations with $f(R)$ gravity and massive neutrinos and reached conclusions broadly similar to our own \cite{GarciaFarietaetal2019}.

\section*{Acknowledgement}

We would like to thank Benjamin Bose for assistance with \MGcopter{}. BSW is supported by the U.K. Science and Technology Facilities Council (STFC) research studentship. KK and HAW are supported by the European Research Council through 646702 (CosTesGrav). KK is also supported by the UK Science and Technologies Facilities Council grants ST/N000668/1. GBZ is supported by NSFC Grants 1171001024  and  11673025,  and  the  National Key Basic Research and Development Program of China (No.  2018YFA0404503). Numerical computations for this research were done on the Sciama High Performance Compute (HPC) cluster which is supported by the ICG, SEPNet, and the University of Portsmouth.

\bibliographystyle{JHEP}
\bibliography{References}

\end{document}